\documentclass[aps,prb,twocolumn,groupedaddress,showpacs,floatfix]{revtex4-1}

\usepackage{graphicx}
\usepackage{grffile}
\usepackage{subfigure}
\usepackage{amsmath}
\usepackage{amssymb}
\usepackage{bm}

\graphicspath{{./}{figures/}}

\newcommand{\Li}{LiFeAs}

\DeclareMathOperator{\im}{Im}

\begin{document}

\title{Effects of correlation in {\Li}}

\author{Johannes Ferber}
\email{ferber@itp.uni-frankfurt.de}

 \author{Kateryna Foyevtsova}

\author{Roser Valent\'\i}

\author{Harald O. Jeschke}

\affiliation{Institut f\"ur Theoretische Physik, Goethe-Universit\"at Frankfurt, Max-von-Laue-Strasse 1, 60438 Frankfurt/Main, Germany}

\date{\today}

\begin{abstract}
  We discuss the role of electronic correlations in the iron-based
  superconductor {\Li} by studying the effects on band structure, mass
  enhancements, and Fermi surface in the framework of density
  functional theory combined with dynamical mean field theory
  calculations. We conclude that {\Li} shows characteristics of a
  moderately correlated metal and that the strength of correlations is
  mainly controlled by the value of the Hund's rule coupling $J$. The
  hole pockets of the Fermi surface show a distinctive change in form
  and size with implications for the nesting properties. Our
  calculations are in good agreement with recent angle-resolved
  photoemission spectroscopy and de Haas-van Alphen experiments.


\end{abstract}

\pacs{71.27.+a,74.20.Pq,74.70.Xa,71.18.+y,71.20.-b}

\maketitle

\section{Introduction}
\label{section:introduction}

High temperature iron-based superconductors have been intensively
studied since their discovery four years ago.~\cite{Johnston2010}
Among the various known iron pnictide classes, the 111 family
comprising LiFeAs and LiFeP shows specially interesting features
compared to the other families.  Whereas superconductivity in many
iron pnictide compounds develops in the vicinity of a
spin-density-wave (SDW) state upon doping or application of external
pressure, {\Li} and LiFeP (and LaFePO from the 1111 family) are
non-magnetic and superconductivity evolves without additional doping
or applied pressure.  Of special relevance is {\Li} where $T_c\approx
18$~K~\cite{Tapp2008, Pitcher2008} compared to $T_c\approx
6$~K~\cite{Deng2009} in LiFeP and $T_c\approx 4$~K~\cite{Kamihara2006}
in LaFePO.  In the 1111 and 122 family compounds (with LaFeAsO and
BaFe$_2$As$_2$ as typical examples), the SDW order is generally
attributed to sizable nesting of the electron and hole Fermi
pockets.~\cite{Singh2009} For {\Li} the situation is not quite as
clear: band structure calculations using density functional theory
(DFT) predict an antiferromagnetic ground state with stripelike order
as in the other pnictides, albeit in a shallow energy minimum compared
to the non-magnetic state.~\cite{Singh2008,Zhang2010} In contrast,
angle-resolved photoemission spectroscopy (ARPES) measurements report
only poor nesting.~\cite{Borisenko2010} In fact, recent neutron
scattering measurements find strong SDW
fluctuations~\cite{Taylor2011,Qureshi2011} with an incommensurate
vector~\cite{Qureshi2011} slightly shifted from the commensurate order
observed in the other iron pnictide superconductors. Also functional
renormalization group calculations\cite{Platt2011} predict SDW fluctuations to be the dominant
instability. On the other side, recent de Haas-van Alphen (dHvA) experiments claim to be in good
agreement with DFT regarding the topology of the Fermi
surface.~\cite{Putzke2011}

It is of particular interest to identify the role of electronic
correlations in this context.  Starting from band structure
calculations within DFT, we include correlations at the level of the
dynamical mean field theory (DMFT) and analyse their effect on the
electronic structure of {\Li}. The band structure of {\Li} features
two shallow hole pockets around the Gamma point which generate a large
density of states and it has been suggested that this is essential for
the way superconductivity emerges in this
compound.~\cite{Borisenko2010,Brydon2011} These features of the
electronic structure can also be expected to be rather susceptible to
changes induced by correlations. Thus, this paper aims to single out
the effects of correlations on the Fermi surface and the low-energy
properties of {\Li}.

\section{Methods and interaction parameters}
\label{section:methods}

Our calculations were performed using an LDA+DMFT implementation which
combines electronic structure calculations in the full potential
linearized augmented plane wave (FLAPW) framework with
DMFT.~\cite{Aichhorn2009} The electronic structure calculations were
executed in WIEN2k~\cite{Blaha2001}, where the self-consistency cycle
employed 1080 $k$ points in the irreducible Brillouin zone, using the
local density approximation~\cite{Perdew1992} (LDA) for the
exchange-correlation potential. We base our calculations on the
experimental crystal structure as obtained from X-ray diffraction
data~\cite{Morozov2010} with space group $P\,4/nmm$. For comparison, we also
performed calculations on the structure given in Ref.~\onlinecite{Tapp2008} for which
we gave mass enhancements in Tab.~\ref{tab:meff}.

For the construction of localized Wannier-like orbitals for DMFT, an
energy window ranging from -5.5~eV to 2.85~eV was chosen, comprising
the Fe $3d$ and As $4p$ bands. For the solution of the DMFT impurity
problem, we employ paramagnetic calculations with the strong-coupling
continuous-time quantum Monte Carlo method~\cite{Werner2006} as
implemented in the ALPS code~\cite{Bauer2011, Gull2011} and consider only the
density-density terms of the Hund's coupling; we used $1 \times 10^7$
Monte Carlo sweeps throughout our calculations.

For the interaction parameters, we use the definitions of $U=F^0$ and
$J=(F^2+F^4)/14$ in terms of Slater integrals~\cite{Anisimov1997} $F^k$, and
the fully localized limit (FLL) double counting
correction.~\cite{Anisimov1993,Dudarev1998}

There is considerable disagreement in the literature about the size of
the interaction parameters in the iron pnictides; in particular the
Coulomb interaction $U$ strongly depends on the estimation method,
whereas $J$ is only moderately reduced from its atomic
value. Self-consistent GW determination yields rather large numbers
(e.g. $U=4.9$~eV, $J=0.76$~eV for BaFe$_2$As$_2$~\cite{Kutepov2010}),
with lower values being reported by constrained LDA (e.g. $U=3.1$ eV,
$J=0.91$~eV for LaFeAsO~\cite{Kutepov2010}) and constrained
random-phase approximation (cRPA) (e.g. $U=2.9$ eV, $J=0.79$~eV for
LaFeAsO~\cite{Aichhorn2009}). For {\Li}, interaction parameters
obtained from cRPA have been reported in Ref.~\onlinecite{Miyake2010}
for two low-energy models, one constructed for the Fe $3d$ bands only,
the other one for a manifold containing Fe $3d$ and As $4p$
states. The choice of the model affects the value of the interaction
parameters in two ways: a model with more bands renders the associated
Wannier functions more localized and thereby increases the matrix
elements of the interaction. Also, since the interaction strength is
derived as a partially screened Coulomb interaction where screening
channels within the low-energy space are subtracted, the exclusion of
more screening channels in a model with more bands increases the
interactions. This is reflected by very different interaction
parameters for the two models, $U=2.45$~eV and $J=0.61$~eV for the $d$
model, $U=4.95$~eV and $J=0.87$~eV for the $dp$ model.

However, as pointed out in Ref.~\onlinecite{Miyake2008}, the
appropriate model for our LDA+DMFT approach is a hybrid model where
the Wannier functions are constructed from a $dp$ model but only
$d$--$d$ transitions are excluded from the screening since we only
treat the $d$ states as correlated in our DMFT procedure. This means
that the $d$ model systematically underestimates the interactions for
our setup whereas the $dp$ model systematically overestimates them. In
light of these uncertainties we report in the following results for
$U=4$~eV, $J=0.8$~eV and include a discussion about the sensitivity of
our results to the choice of interaction parameters in
Sec.~\ref{subsection:sensitivity}.

In order to obtain real-frequency spectra from the imaginary time
Monte Carlo data we performed analytic continuation of the self energy
using the classic Maximum Entropy method.~\cite{Jarrell1996} To avoid
uncertainties from the analytic continuation, the effective masses and
Fermi surfaces are directly infered from the self energy on the
Matsubara axis: the mass enhancements read
\begin{equation}
 m^\ast/m_{\textrm{LDA}} = 1- \left.\frac{\partial \im \Sigma(\textrm{i}\omega)}{\partial
\omega}\right|_{\omega
\rightarrow 0^+} \; ,
\label{eq:m_eff}
\end{equation}
where the derivative is extracted by fitting a fourth-order polynomial
to the data for the lowest six Matsubara
frequencies.~\cite{Mravlje2011} The same polynomial is used for the
determination of the Fermi surfaces where we make use of the fact that
the imaginary and real axis meet at zero,
$\Sigma(\omega=0)=\Sigma(i\omega=0)$.

Calculations were performed at a temperature $T=72.5$~K
($\beta=160$~eV$^{-1}$).

In the following, orbital characters are labelled in a coordinate
system which is 45$^{\circ}$ rotated with respect to the
crystallographic axes, {\it i.e.} $x$ and $y$ point to nearest Fe
neighbors in the Fe-As plane.

\section{Results}
\label{section:results}

\begin{figure}[!h]
\includegraphics[width=\columnwidth]{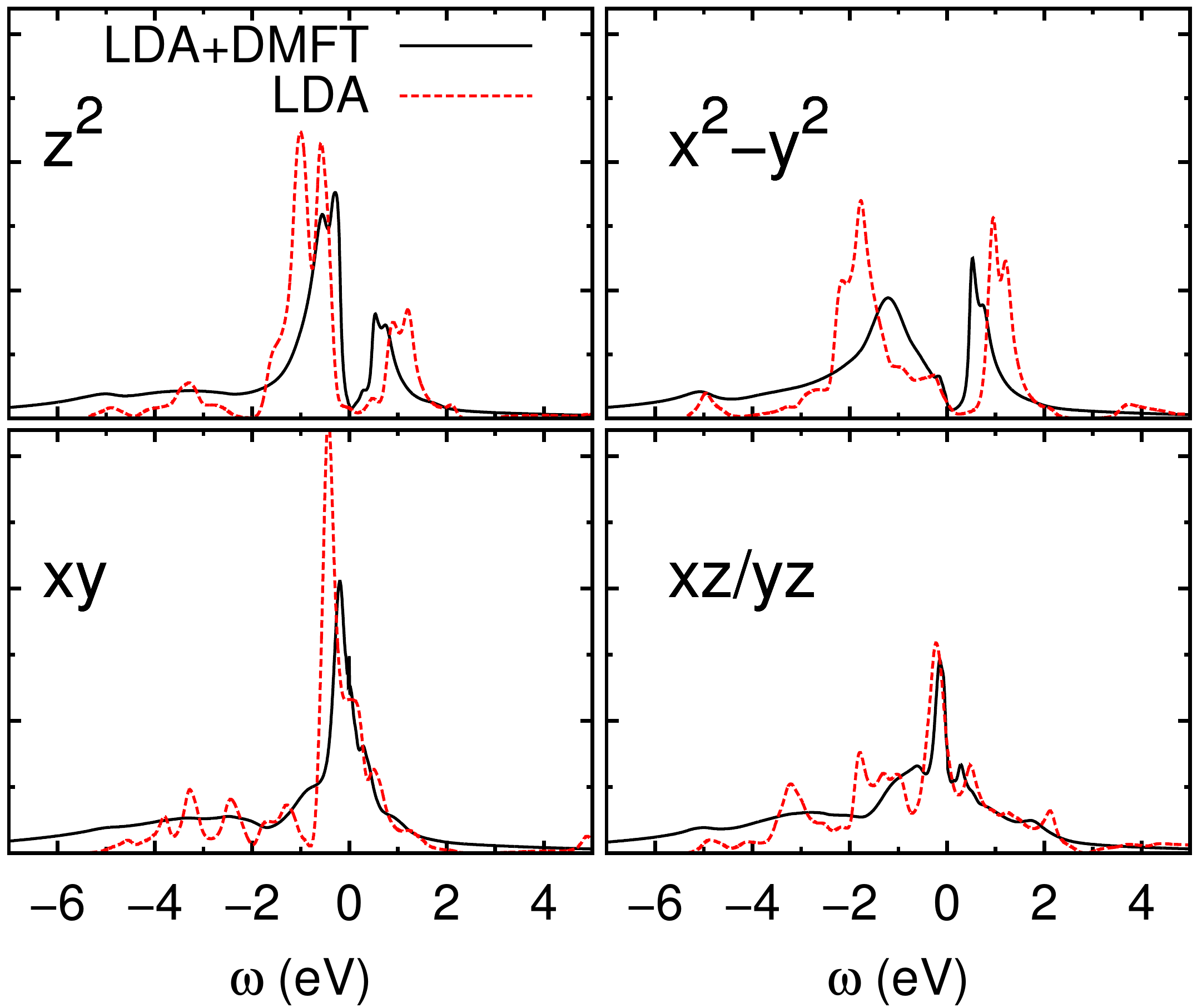}
\caption{\label{fig:A}(Color online) Orbital-resolved comparison
  between LDA density of states (red dotted lines) and the LDA+DMFT spectral
  function $A(\omega)$ (black full lines). The interaction parameters
used were $U=4$~eV, $J=0.8$~eV.}
\end{figure}

\begin{figure}[!h]
\includegraphics[width=\columnwidth]{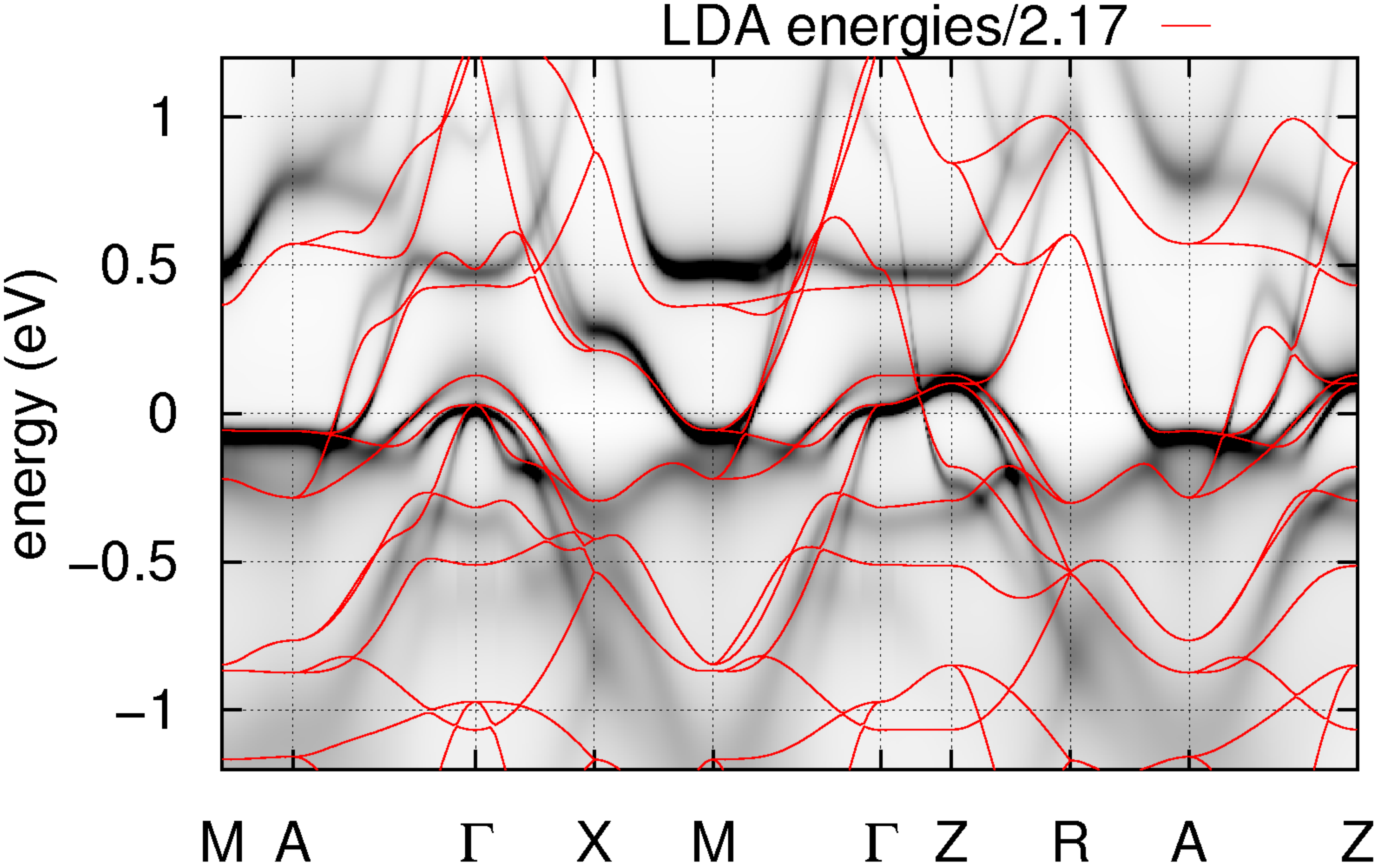}
\caption{\label{fig:Ak}(Color online) Momentum-resolved spectral
  function  $A(\bf{k},\omega)$ together with LDA bands.  For comparison, the LDA band
  energies are divided by the orbitally averaged value of the mass
  renormalization.  The interaction parameters are the
same as in Fig.~\ref{fig:A}.}
\end{figure}

In Figs.~\ref{fig:A} and \ref{fig:Ak}, we compare the
momentum-integrated and momentum-resolved spectral function for LiFeAs
obtained within LDA+DMFT with their LDA counterparts, namely the
density of states (Fig.~\ref{fig:A}) and the LDA band energies
(Fig.~\ref{fig:Ak}). Note that the LDA bands in Fig.~\ref{fig:Ak} were
renormalized by a factor of 2.17 corresponding to the orbitally
averaged value of the mass renormalization.

The momentum-integrated spectral function $A(\omega)$ shows a
bandwidth reduction but no substantial spectral weight transfer, {\it
  i.e.} no formation of Hubbard bands. The momentum-resolved spectral
function $A(\bf{k},\omega)$ in Fig.~\ref{fig:Ak} displays well-defined
excitations around the Fermi level and stronger correlation-induced
broadening of the states at higher binding energies. The broadening
affects the states below the Fermi level more strongly where coherent
quasiparticles can be identified down to approx. 0.3~eV below $E_F$. For
the states above $E_F$, the crossover to rather diffuse structures
occurs at approx. 0.7~eV. On a quantitative level, at our temperature
$T=72.5$~K, the scattering rates (or, equivalently, inverse lifetimes)
-Im$\Sigma(i0^+)$ are small, see Tab.~\ref{tab:meff}, supporting the
picture of well-defined, long-lived quasiparticles. The renormalized
LDA bands give a good approximation only close to the Fermi level (the
mass enhancement in Eq.~(\ref{eq:m_eff}) holds strictly only at
$\omega=0$); states away from $E_F$ are less renormalized.

For the given interaction parameters, the self energy and spectral
function thus show the characteristics of a Fermi liquid state in a
metal with moderate correlations, a picture which also has been
promoted for the 1111 and 122 family of iron pnictides in a number of
previous
publications.~\cite{Anisimov2009,Aichhorn2009,Skornyakov2009,Skornyakov2010,Hansmann2010}
Note that for multiorbital systems with sufficiently strong $J$, the
absence of the rotationally invariant Hund's coupling in the calculation
({\it i.e.} the consideration of the density-density terms in $J$
only) can lead to qualitatively wrong results by suppressing coherence
and driving the system from a Fermi liquid into a non-Fermi liquid
state.~\cite{Haule2009,Ishida2010,Aichhorn2010} This is not observed
here, indicating that the restriction to density-density terms in the
Hund's coupling is not detrimental. We also estimated the temperature
below which coherent quasiparticles form by calculating
$\chi(\tau=\beta/2)$, {\it i.e.} the paramagnetic local spin
susceptibility at imaginary time point $\beta/2$. In a Fermi liquid
$\chi(\tau=\beta/2)\sim T^2$, and by studying the temperature
dependence of $\beta^2 \chi(\tau=\beta/2)$ one finds that for $U=4$~eV
and $J=0.8$~eV it takes on a constant value only at low temperatures
below $\approx 100$~K.

The mass enhancements as given in Tab.~\ref{tab:meff} exhibit
pronounced orbital dependence, with stronger mass enhancement in the
$t_{2g}$ orbitals $d_{xy}$ and $d_{xz}$/$d_{yz}$. As can also be seen
from Fig.~\ref{fig:A}, the bandwidth $W$ of the $t_{2g}$ orbitals is
smaller, leading to a larger ratio $U/W$ and to increased correlations
in these orbitals.  Analysis of a low-energy iron $d$ tight-binding
model obtained by considering the localized Wannier orbitals shows
that the diagonal nearest neighbor hopping for the $d_{xy}$ orbital,
$t_{\textrm{NN}}(xy,xy)$, almost vanishes as the direct hopping from
the iron-iron overlap and the indirect hopping from the
iron-pnictogen-iron overlap have opposite signs and almost
cancel. Also the diagonal hoppings to further iron neighbors for
$d_{xy}$ are small; this contributes to the localization of the
$d_{xy}$ quasiparticles and a stronger mass enhancement than in the
other orbitals.~\cite{Yin2011} The table lists the mass enhancements for both investigated
structures which show some quantitative yet not qualitative differences.
We checked that in particular Fermi surfaces are practically not affected, though; we
therefore continue to give results only for the structure from Ref.~\onlinecite{Morozov2010}.

\begin{table}[!h]
  \caption{\label{tab:meff} Orbital-resolved quasiparticle weights
    $Z$, mass enhancements $m^\ast/m_{\rm LDA}$, and scattering rates
    -Im$\Sigma(i0^+)$ for interaction parameters $U=4$~eV, $J=0.8$
    eV. The first (second) number in each cell refers to calculations performed on the structure
from Ref.~\onlinecite{Morozov2010} (Ref.~\onlinecite{Tapp2008}).}
\begin{ruledtabular}
\begin{tabular}{lrrrr}
Orbital & $d_{z^2}$ & $d_{x^2-y^2}$ & $d_{xy}$ & $d_{xz/yz}$ \\
\hline
$Z$ & 0.57/0.53 & 0.64/0.60 & 0.36/0.31 & 0.42/0.36 \\
$m^\ast/m_{\rm LDA}$ & 1.74/1.88 & 1.57/1.67 & 2.78/3.24 & 2.39/2.78 \\
-Im$\Sigma(i0^+)$ (meV) & 0.1/0.3 & -1.0/-0.7 & 2.4/5.2 & 1.7/3.8
\end{tabular}
\end{ruledtabular}
\end{table}

\begin{figure}[!h]
\includegraphics[width=\columnwidth]{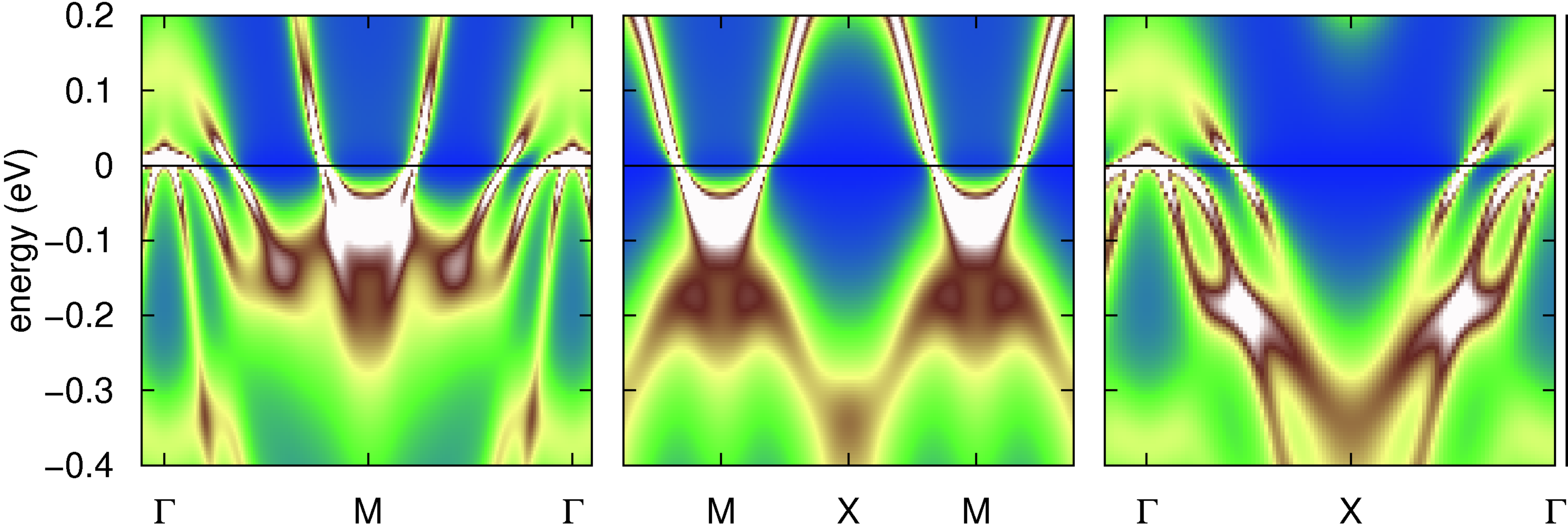}
\caption{\label{fig:Ak_cuts_borisenko}(Color online) LDA+DMFT
  momentum-resolved spectral function $A(\bm{k},\omega)$ for LiFeAs
  along the paths in the Brillouin zone given in
  Ref.~\onlinecite{Borisenko2010}, Fig.~2e. Interaction parameters as
  in Fig.~\ref{fig:A}.}
\end{figure}

For a comparison with ARPES measurements,
Fig.~\ref{fig:Ak_cuts_borisenko} shows some cuts of the
momentum-resolved spectral function $A(\bm{k},\omega)$ along the paths
given in Fig.~2e in Ref.~\onlinecite{Borisenko2010}.  Qualitatively,
we find good agreement; quantitatively, the mass enhancement extracted
from these cuts in Ref.~\onlinecite{Borisenko2010} is 3.1. This value
should be compared to the mass enhancements of the orbitals that
contribute most to the spectral weight at low energy. As can be seen
from Fig.  \ref{fig:A}, these are the $t_{2g}$ orbitals (the $e_g$
orbitals show a dip around the Fermi level) with calculated mass
enhancements of 2.4--2.8 (2.8--3.2, respectively). Thus, mass enhancements are in good
agreement, with ARPES pointing to moderately larger
interactions. We will come back to this point further below.

In summary we consider {\Li} a metal in an intermediate range of interactions without significant
spectral weight transfer. Mass renormalizations are close to what has been calculated and measured
in the 1111 and 122 systems; compared to e.g. LaFeAsO~\cite{Aichhorn2011}, coherent quasiparticles
seem to form a lower temperatures, though, with the spin susceptibility approaching Fermi
liquid-like behavior only below $\approx 100$~K.

\begin{figure}[!h]
\includegraphics[width=\columnwidth]{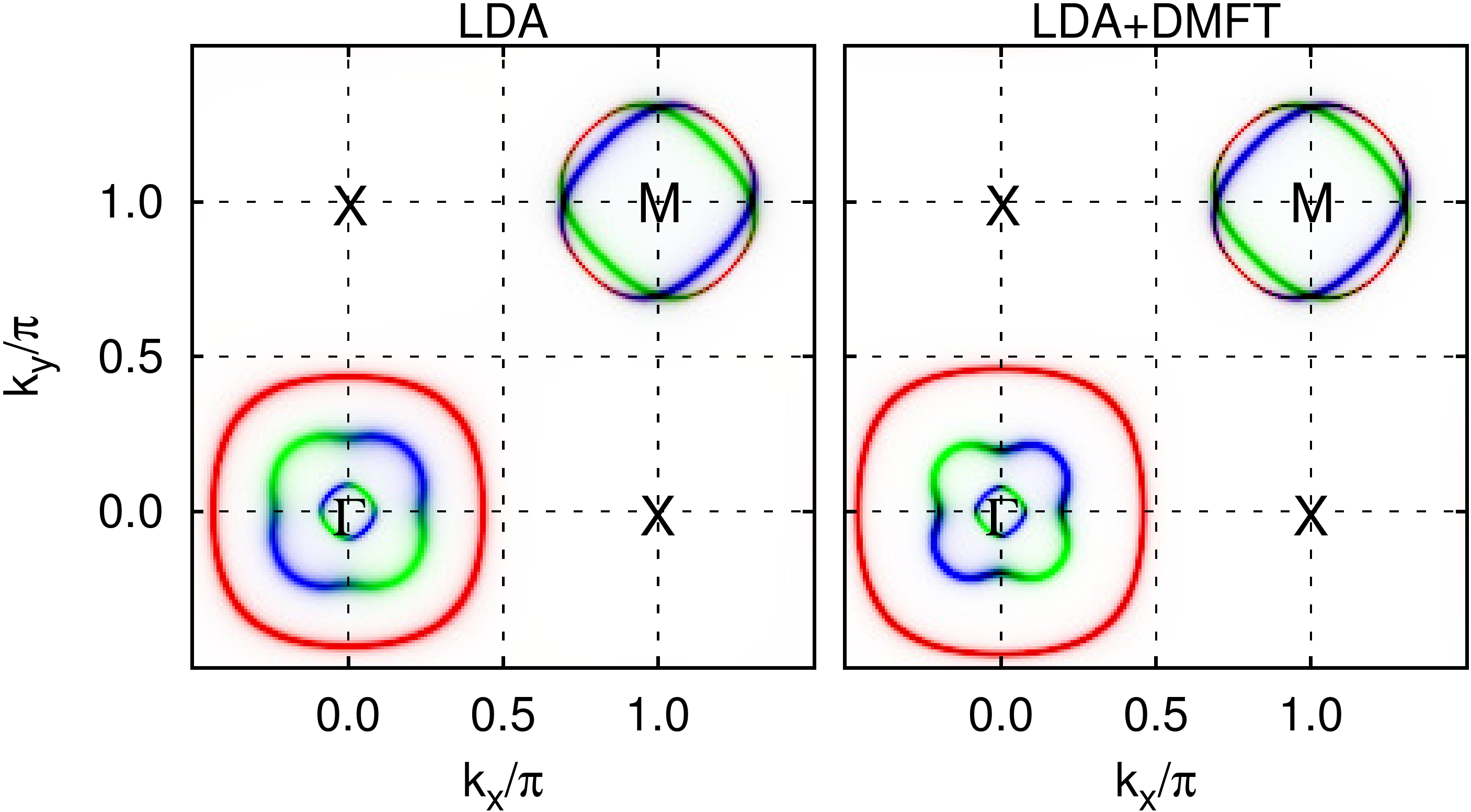}
\caption{\label{fig:fermi_kz0}(Color online) Fermi surface for
  $k_z=0$. Color code for orbital characters: red: $d_{xy}$, green:
  $d_{xz}$, blue: $d_{yz}$. Interaction parameters as in
  Fig.~\ref{fig:A}.}
\end{figure}

\begin{figure}[!h]
\includegraphics[width=\columnwidth]{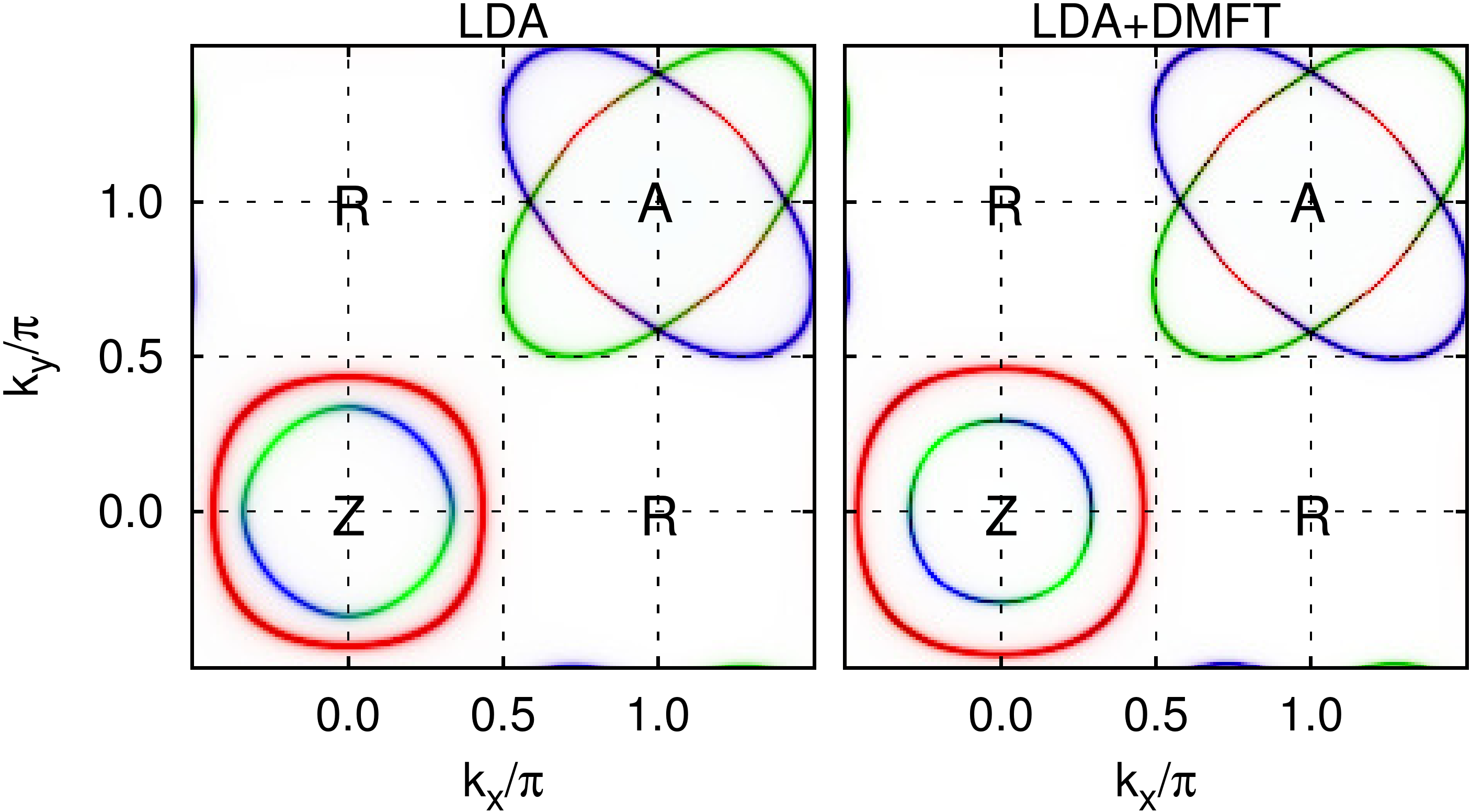}
\caption{\label{fig:fermi_kz05}(Color online) Fermi surface for
  $k_z=\pi$. Color code and interaction parameters as in
  Fig.~\ref{fig:fermi_kz0}.}
\end{figure}

We now turn our attention to the discussion of the effects of
correlations on the Fermi surfaces of {\Li} which have been
experimentally accessed by ARPES and dHvA~\cite{Putzke2011}
measurements.  Figs.~\ref{fig:fermi_kz0} and \ref{fig:fermi_kz05} show
the Fermi surfaces in the $k_z=0$ and $k_z=\pi$ plane obtained within
LDA and LDA+DMFT. The pockets around $(k_x,k_y)=(0,0)$ are hole
pockets while the ones around $(k_x,k_y)=(\pi,\pi)$ are electron
pockets (compare Fig.~\ref{fig:Ak}). The most prominent effects of
correlations are the shrinking of the middle $d_{xz}$/$d_{yz}$ hole
pocket which takes on a butterfly shape at $k_z=0$, and the increase
of the outer $d_{xy}$ pocket whereas the electron pockets almost don't
change in size or form. This observation is in agreement with previous
calculations~\cite{Yin2011} and would support ARPES results. This
analysis shows that correlations tend to weaken --if not suppress--
nesting in this material.

For the discussion of the electron pockets, we describe the Fermi
surface in terms of an inner and outer pocket rather than by two
crossed ellipses-like pockets of equal size. This is motivated by the
fact that spin-orbit (SO) coupling lifts the degeneracy between the
ellipses and splits the electron pockets into an inner and outer
sheet.~\cite{Putzke2011,relativistic,Note2} Note, however, that no SO
coupling is taken into account in the present calculation. As one can see from
the comparison of Fig.~\ref{fig:fermi_kz0} and
Fig.~\ref{fig:fermi_kz05}, the thus defined outer pocket has strong
$k_z$ dispersion whereas the inner sheet depends only weakly on $k_z$.

\begin{figure}[!h]
\includegraphics[width=\columnwidth]{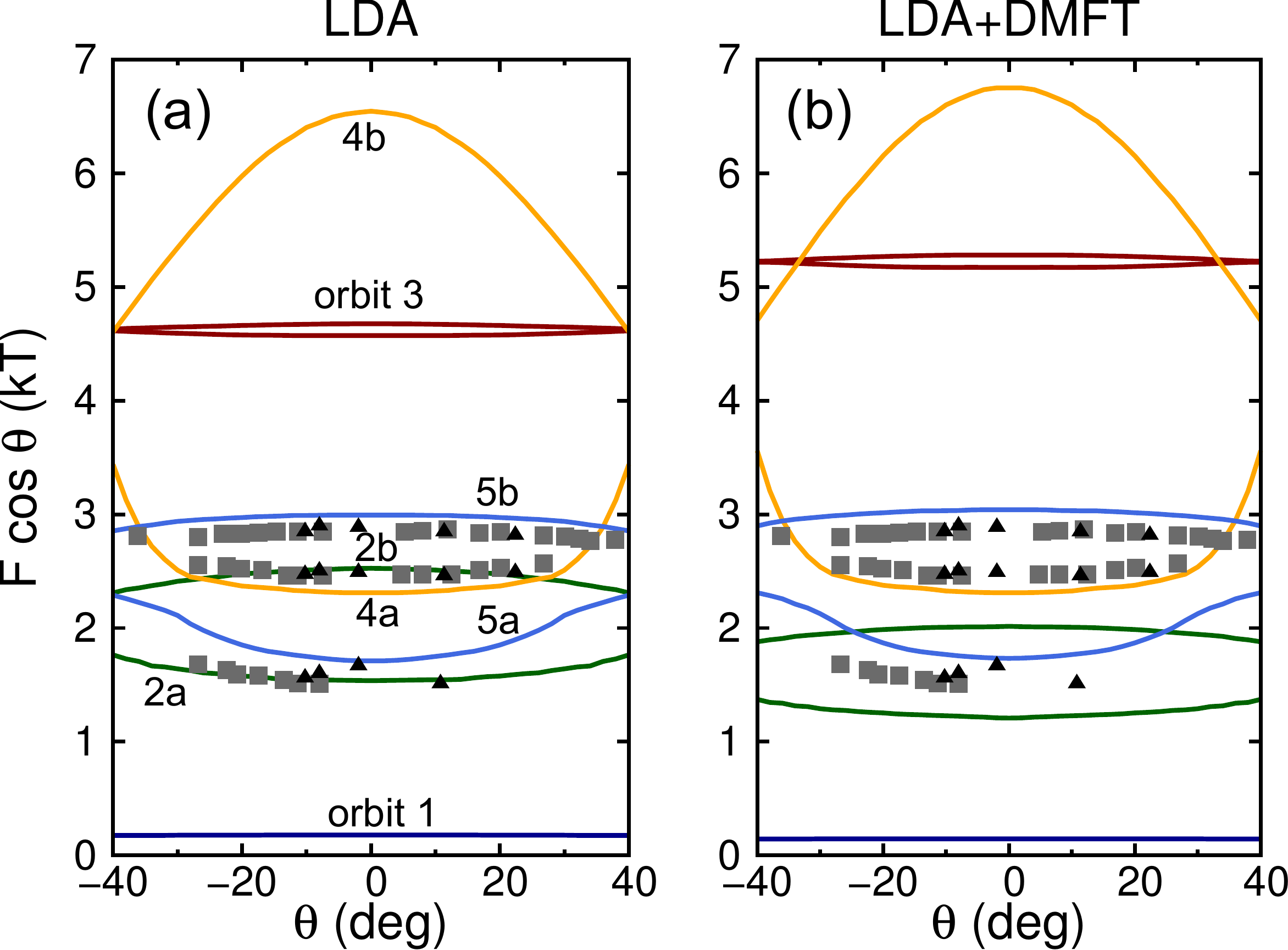}
\caption{\label{fig:dhva}(Color online) dHvA frequencies with respect
  to magnetic field angle obtained within (a) LDA and (b) LDA+DMFT.
    The orbits refer to extremal pocket sizes
  where the pockets are identified as discussed in the
  text. Interaction parameters as in Fig.~\ref{fig:A}. Triangles (pulsed field) and squares (dc
field) are experimental data taken from Ref.~\onlinecite{Putzke2011}.}
\end{figure}

In order to facilitate a quantitative comparison with the dHvA
measurements, we show in Fig. \ref{fig:dhva} calculated dHvA
frequencies with respect to magnetic field angle as reported in
Fig.~2c of Ref.~\onlinecite{Putzke2011}. The dHvA frequencies
correspond to extremal pocket sizes (orbits) that are observed at a
given angle $\theta$ with respect to the $k_z$ axis. The labelling of
the orbits follows Ref.~\onlinecite{Putzke2011}: orbits 1, 2, and 3
refer to the inner, middle, and outer hole pocket, and orbits 4 and 5 to
the outer and inner electron pocket (see Fig.~\ref{fig:dhva} (a)).
 In order to define pocket sizes
within LDA+DMFT  (Fig. \ref{fig:dhva} (b))  in view of the finite broadening induced by the
correlations, we track the maximum of $A(\omega=0)$ through
the Brillouin zone.

Compared to the calculated dHvA frequencies in
Ref.~\onlinecite{Putzke2011} small differences are already
visible on the LDA level (Fig.~\ref{fig:dhva} (a)), e.g. the minima of orbit 2 and orbit 5 for
small angles are not degenerate anymore. This is probably an effect of
differences in the determination of the Fermi surface (e.g. due to effects of a finite
k-mesh) and illustrates the high sensitivity of the
orbits to to details of the calculation.

As already seen in Figs.~\ref{fig:fermi_kz0} and \ref{fig:fermi_kz05},
the effect of correlations on the Fermi surface manifests itself
mainly in a shrinking of the middle hole pocket, and, in order to
preserve the electron count, an increase of the outer hole pocket
size. This is reflected merely by a shift downwards of orbit 2 and
a shift upwards of orbit 3 in 
Fig.~\ref{fig:dhva} (b);
a change in the warping would indicate a change
in the $k_z$ structure of the pocket which is not to be expected from
the local, {\it i.e.}  $k$-independent interaction in (single-site)
DMFT.

Analyzing the curvature and the size of the orbits, the authors of
Ref.~\onlinecite{Putzke2011} attribute the experimentally measured
frequencies to the electron Fermi surface sheets, where the two higher
frequencies are assigned to orbits 5b and 4a and the lowest frequency
is suspected to originate from orbit 5a.~\cite{Note2} Our results
support this interpretation: whereas the orbits 2a/5a and 2b/4a are of
similar size in the LDA calculation, the correlations affect mainly
the hole pockets and lift this (near-)degeneracy. As a result, the
electron orbits 2a and 2b are unlikely to give rise to the measured
frequencies as their sizes are rather different from the measured
data. This offers a reconciliation of the dHvA and ARPES experiments:
the shrunk middle hole pocket is only seen in ARPES which finds a
correlated metal with poor nesting together with sizable mass
renormalization. In contrast, the dHvA measurements resolve  the
 (lighter) electron pocket sizes in {\Li} that almost do not change under
inclusion of correlation and therefore report good agreement with
LDA. The large mass renormalizations (up to $\approx$5) that
are also measured in Ref.~\onlinecite{Putzke2011} suggest --even under
consideration of a non-negligible electron-phonon contribution-- a
scenario of important electronic correlations which are correctly
accounted for within LDA+DMFT.

In LiFeP, which was also investigated in the same work Ref.~\onlinecite{Putzke2011}, the situation
is
different. Whereas the middle hole pocket in {\Li} is particulary shallow with a band energy
of 64 meV above $E_F$ at the $\Gamma$ point, the pocket in LiFeP is roughly of the same size as in
{\Li}, however the corresponding band energy at $\Gamma$ is 155 meV. Hence, the
dispersion in LiFeP is considerably stronger, rendering the pocket less susceptible to band shifts
as induced by the real part of the selfenergy, i.e. less susceptible to correlations. Consequently,
the dHvA frequencies remain rather unchanged upon inclusion of correlations.

\subsection*{Sensitivity to interaction parameters}
\label{subsection:sensitivity}

We analyze the sensitivity of our results to our choice of interaction
parameters by applying some variation to $U$ and $J$ while keeping the
respective other parameter fixed.

\begin{figure}[!h]
\includegraphics[width=\columnwidth]{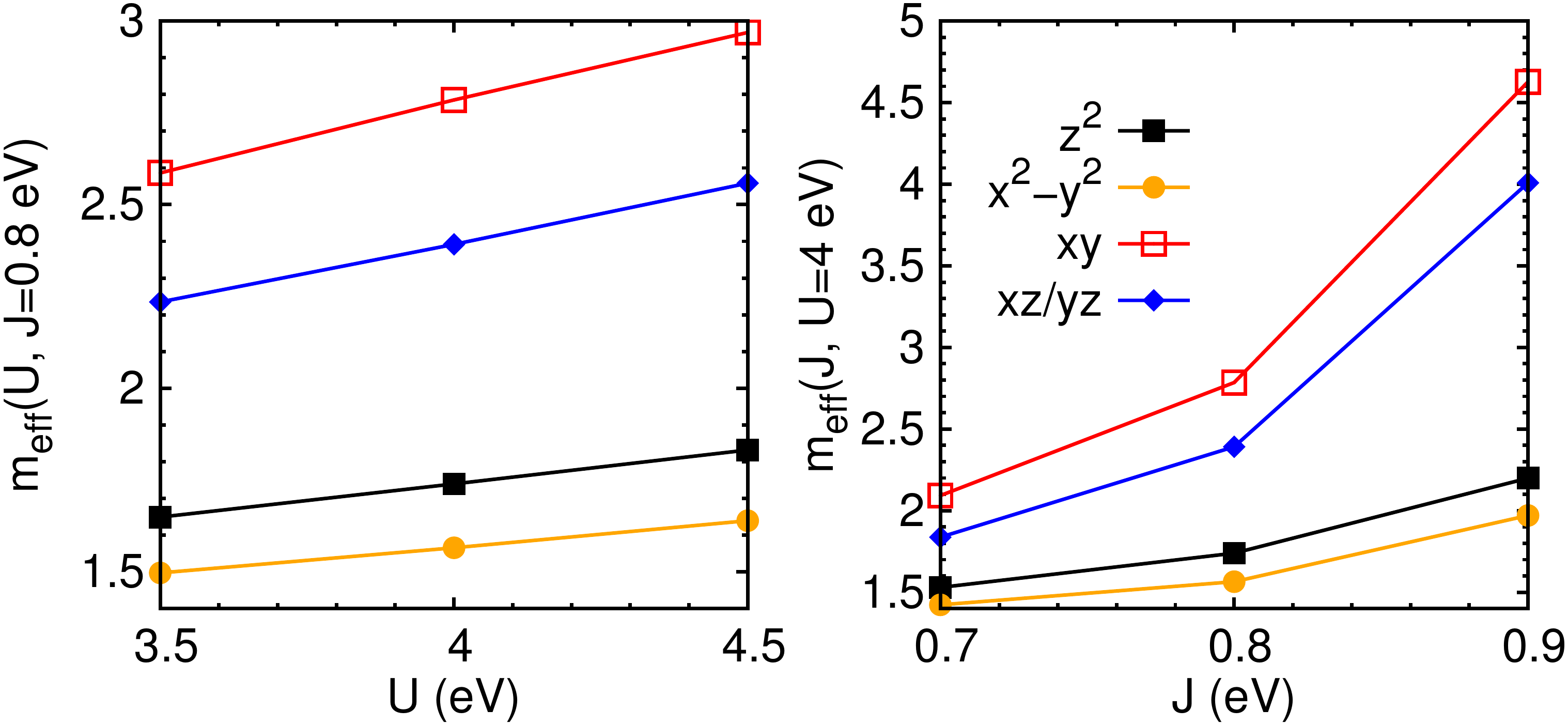}
\caption{\label{fig:sensitivity_meff}(Color online) Sensitivity of
  effective masses  $m^\ast/m_{\textrm{LDA}}$  with respect to changes in the interaction
  parameters.}
\end{figure}

Fig.~\ref{fig:sensitivity_meff} shows the evolution of the mass
enhancements $m^\ast/m_{\textrm{LDA}}$ with the interaction parameters.
 A moderate dependence on
$U$ and a very strong dependence on $J$ are observed (note that the
applied variations of $U$ and $J$ are different in size). Whereas a
change in $U$ affects all orbitals roughly equally, a change in $J$
leads to an immense mass enhancement particulary of the $t_{2g}$
orbitals.

The decisive role of the Hund's coupling for the physical
properties of the iron pnictides has been discussed 
previously.~\cite{Haule2009} For the different behavior of the
$e_g$ and $t_{2g}$ orbitals it is important that the $e_g$ states in {\Li} lie
energetically lower than the $t_{2g}$ states. In the atomic limit, the
energy gain from Hund's rule exceeds the crystal field splitting
already for rather small $J$ and the ground state is a high spin state
with the configuration $e_g^3t_{2g}^3$ where the $t_{2g}$ orbitals are
occupied by three electrons of the same spin. In the atomic limit this
prevents mixing of the orbitals due to the Pauli principle; in the crystal, it still impedes
inter-orbital fluctuations within the $t_{2g}$
manifold.~\cite{Medici2011,Yin2011} This effect contributes to the
high sensitivity of the $t_{2g}$ effective masses with respect to $J$.

\begin{figure}[!h]
\includegraphics[width=\columnwidth]{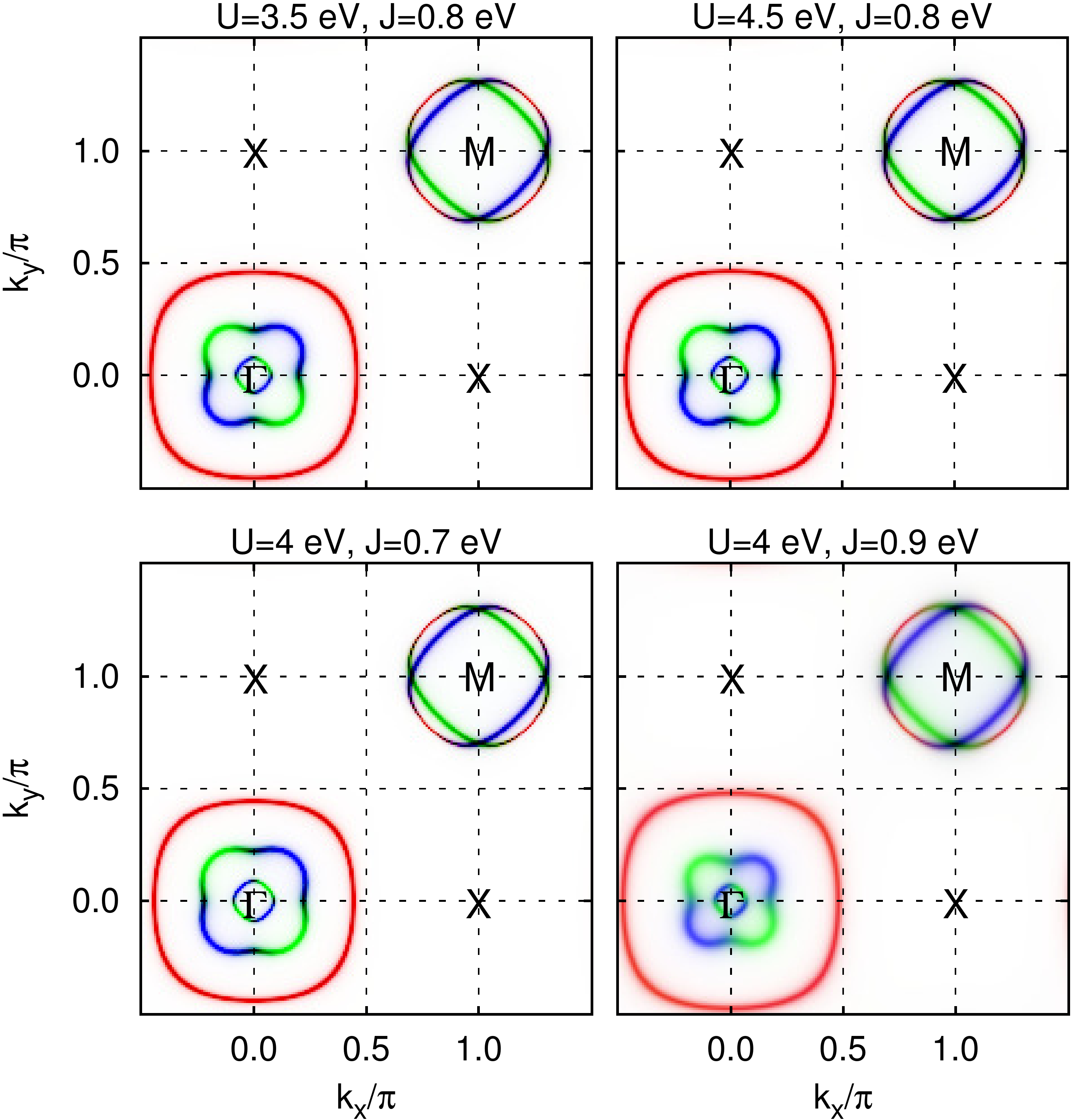}
\caption{\label{fig:sensitivity_fermi_kz0}(Color online) Sensitivity
  of the Fermi surface at $k_z=0$ with respect to changes in the
  interaction parameters.}
\end{figure}

\begin{figure}[!h]
\includegraphics[width=\columnwidth]{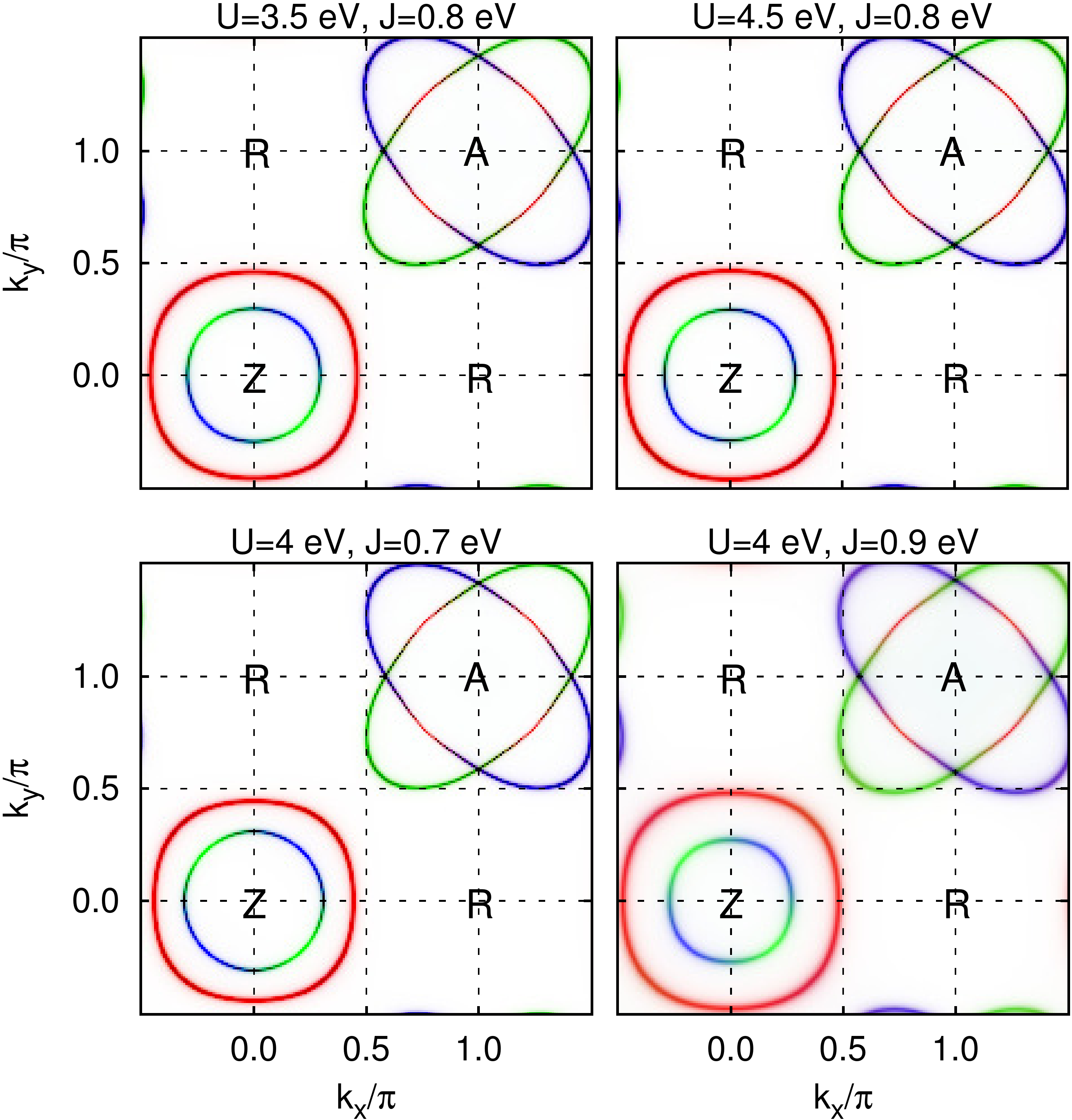}
\caption{\label{fig:sensitivity_fermi_kz05}(Color online) Sensitivity
  of the Fermi surface at $k_z=\pi$ with respect to changes in the
  interaction parameters.}
\end{figure}

Accordingly, the Fermi surface is rather stable against variation of
$U$ but strongly depends on $J$, shown for $k_z=0$ in
Fig.~\ref{fig:sensitivity_fermi_kz0} and $k_z=\pi$ in
Fig.~\ref{fig:sensitivity_fermi_kz05}. Following the trend discussed
above, larger values of $J$ promote a more pronounced shrinking
(increase) of the middle (outer) hole pocket. Values as large as 0.9
eV for the Hund's coupling render the system rather incoherent though,
with significant scattering rates -Im$\Sigma(i0^+)$ around 14 meV on
the $t_{2g}$ orbitals. This causes the broadening of the respective
Fermi surface in Figs.~\ref{fig:sensitivity_fermi_kz0} and
\ref{fig:sensitivity_fermi_kz05}. Future calculations with the full
Hund's rule coupling are required to check
whether the occurrence of the coherence-incoherence crossover
already  at $J\lesssim 0.9$~eV is 
 a physical effect
or an artifact of the breaking of the rotational invariance by
density-density interactions.

\section{Conclusions}
\label{section:conclusions}

We showed that within the considered range of interaction parameters
{\Li}  behaves as a Fermi liquid where correlation
effects are very sensitive to the value of the Hund's rule coupling in
particular for the $t_{2g}$ orbitals.
The  strong mass enhancements measured in both ARPES and dHvA experiments suggest
sizable correlations of the size considered in this work. While electron-phonon effects
have been reported~\cite{Kordyuk2011} to be significant and contribute to the slightly higher
mass enhancements measured experimentally, they cannot account alone for the large values observed.
As for the Fermi surface, the correlations mainly affect the hole pockets that significantly change
in size. We propose this as the source of the seeming discrepancy of the ARPES and dHvA
experiments: whereas dHvA presumably observes only electron orbits with sizes close to their LDA
values, ARPES finds the reduced size of the middle hole pocket as most prominent feature. In this
way, the two experiments can be reconciled. The selective size reduction of the middle
hole pocket also renders nesting less efficient.

\medskip

{\it Acknowledgements}.- We would like to thank C. Putzke, 
B. B\"uchner, and especially  A. Coldea and M.  Aichhorn for
very  useful discussions. We
gratefully acknowledge financial support from the Deutsche
Forschungsgemeinschaft through the SPP 1458 program and from the
Helmholtz Association through grant HA216/EMMI.

\appendix


\begin{thebibliography}{37}
\bibitem{relativistic} {We performed fully relativistic
    calculations~\cite{FPLO} of the Fermi surface of {\Li} to confirm
    the avoided crossing of the electron sheets.}
\end{thebibliography}
\end{document}